\documentclass[prl, twocolumn, letterpaper, showpacs]{revtex4}
\usepackage{graphicx}
\usepackage{amsmath}
\usepackage{ulem}
\usepackage[english]{babel}
\usepackage[hypertex]{hyperref}
\hypersetup{
    colorlinks,
    citecolor=red,
    filecolor=blue,
    linkcolor=blue,
    urlcolor=blue
}

\begin{document}

\title{Rabi oscillations, decoherence, and disentanglement in a qubit-spin-bath system: exact dynamics}

\author{Ning Wu\footnote{These two authors equally contributed to the work.}, Arun Nanduri$^*$, and Herschel Rabitz\footnote{hrabitz@princeton.edu}}

\affiliation{\it Department of Chemistry, Princeton University, Princeton, NJ 08544}


\begin{abstract}
 We examine the influence of environmental interactions on simple quantum systems by obtaining the exact reduced dynamics of a qubit coupled to a one-dimensional spin bath. In contrast to previous studies, both the qubit-bath coupling and the nearest neighbor intrabath couplings are taken as the spin-flip XX-type. We first study the Rabi oscillations of a single qubit with the spin bath prepared in a spin coherent state, finding that nonresonance and finite intrabath interactions have significant effects on the qubit dynamics. Next, we discuss the bath-induced decoherence of the qubit when the bath is initially in the ground state, and show that the decoherence properties depend on the internal phases of the spin bath. By considering two independent copies of the qubit-bath system, we finally probe the disentanglement dynamics of two noninteracting entangled qubits. We find that entanglement sudden death appears when the spin bath is in its critical phase. We show that the single-qubit decoherence factor is an upper bound for the two-qubit concurrence.
\end{abstract}

\maketitle

\section{I. Introduction}
The quantum dynamics of a single qubit or central spin coupled to a spin environment~\cite{RPP} has been widely studied theoretically in several different areas, including quantum information sciences~\cite{Bose2004,Fazio2007,Lukin2011,Pla2012,Chekhovich2013,Warburton2013,Nori2013,Imamoglu2013}, quantum decoherence~\cite{Petru2004,Petru2007,Bortz2010,Erbe2010,Sarma2012,Stanek2013,Ratschbacher2013,Wang2013,FaribaultPRL2013,Balian2014}, and excitation energy transfer~\cite{Johnson2008,Ekert2012,Wu2013}. One of the most promising candidates for quantum computation, solid-state spin systems, are inevitably coupled to their surrounding environment, usually through interactions with neighboring nuclear spins~\cite{Chekhovich2013,Dobrovitski2013,Petta2013}. The coupling of a qubit to a spin bath can in general lead to non-Markovian behavior~\cite{Krovi2007,Ferraro2008,Lorenzo2013}, causing the usual Markovian quantum master equations to fail for such models. Most recently, it was demonstrated for the first time that a spin bath can assist coherent transport in a two-level system~\cite{Ekert2012}. Fully understanding the role played by a spin environment is an interesting and important issue.
\par One commonly studied qubit-spin bath system is the so-called spin-star network~\cite{Bose2004,Petru2004,Petru2007,Bortz2010,Stanek2013,Ekert2012,Wu2013,CR2013,FaribaultPRB2013}, in which a preferred central spin is coupled homogeneously to a spin bath without intrabath interactions. A more realistic type of environment takes the form of quantum interacting spin chains~\cite{Fazio2007,Sun2006,Goan2007,Chen2008,Jaksch2010,Lin2011,Busch2012}, where the decay of the qubit's coherence is found to be related to the critical properties of the spin environments. Most prior work making use of such an environment considered qubit-spin bath coupling of the Ising form, which is spin conserving. As a result, it is much easier to analytically obtain the full dynamics of the system, in contrast to the situation where a spin-flip coupling is present. Exceptions include Refs.~\cite{Goan2007,Jaksch2010}, where the authors considered the spin-flip XX-type qubit-bath coupling but with a spin bath having homogeneous self-interactions, and Ref.~\cite{Chen2008}, where the authors use $t$-DMRG to study the reduced dynamics of a qubit coupled locally to an XXZ spin chain via the Heisenberg-type qubit-bath interaction. It should be noted that, in general, both the spin-star network and the homogeneously coupled spin bath can be treated by introducing collective angular momentum operators which facilitates the analytical treatment. In this work, we will focus on a more realistic system with a (not necessarily uniform) spin-flip qubit-bath interaction as well as short range XX-type intrabath interactions. To our knowledge, the exact dynamics of such a model, which is one step closer to faithfully representing environmental spins interacting via fully general Heisenberg-type interactions, has not been obtained before.
\par The collapse and revival (CR) behavior of Rabi oscillations of a qubit coupled to a single bosonic field mode, described by the Jaynes-Cummings (JC) model, is a fundamental consequence of field quantization and provides a much-studied illustration of the quantum nature of qubit-field systems~\cite{CR1990}. Using a correspondence between the JC model and a spin-star network with a large number of spins, it is found in Ref.~\cite{CR2013} that within a certain parameter regime, CR phenomena also appear in a qubit-big spin model. Ref.~\cite{Fehske2012} goes beyond the resonant JC model to the nonresonant Dicke model, and notes that the dynamics depends on the sign of the detuning between the qubit and field frequency. In this work, we extend the model studied in Ref.~\cite{CR2013} to the nonresonant case with a self-interacting spin bath modeled by the periodic XX spin chain. It is found that both nonzero detuning and the nearest neighbor coupling within the XX bath can have an effect on the qubit's dynamics. In particular, the interplay between nonresonance and intrabath interaction is able to reproduce CR behavior even for a spin bath with a relatively small number of sites.
\par In addition, the dynamics of entanglement in many-body systems has recently been studied from different perspectives~\cite{RMP2009}. As interacting quantum spin systems are believed to be paradigmatic for quantum information processing~\cite{Chuang2000}, their entanglement dynamics has attracted much attention~\cite{Amico2004,Sub2004,Sen2005,Jing2007,Furman2008,Bose2009,Wu2010,Tanimura2010,Kais2011,Long2013}. In prior works, the dynamical behavior of pairwise entanglement is found to be related to quantum phase transitions of the spin chains. Another emerging focus is on the evolution of the entanglement of a pair of qubits exposed to noisy environments. In a seminal work, Yu and Eberly~\cite{eberly} found that the Markovian dynamics of the entanglement between two qubits coupled to individual bosonic baths can behave in sharp contrast to single qubit decoherence: the pairwise entanglement of two initially entangled two-level atoms suddenly disappears in a finite time proportional to the spontaneous lifetime of single atoms, while the single atom coherence only vanishes asymptotically. This phenomenon is called entanglement sudden death (ESD). More recently, it has been shown in the same setup that there exists a revival of the vanished entanglement if non-Markovian effects are taken into account~\cite{Compagno2007,Chen2013}. As mentioned earlier, the non-Markovian behavior caused by the spin environment may result in novel dynamics of the pairwise entanglement of two qubits each coupled to their own spin bath, as observed in a locally interacting qubit-spin-bath system~\cite{Chen2008}. ESD and subsequent revivals have also been observed to occur in two qubits when they are coupled to \textit{classical} interacting spin baths~\cite{Zampetaki2012} and external fields~\cite{Franco}, the latter of which has been demonstrated experimentally~\cite{Xu}, and to stochastic noise sources~\cite{Galitski2012}. Furthermore, understanding the relation between decoherence and disentanglement is believed to be of importance both for the foundations of quantum mechanics and practical applications in quantum information science~\cite{eberly,Tolkunov2005,Ann2007,Blatt2010,Kim2012}. In this work, by taking the bath's initial state as the ground state of the XX chain, we first study the decoherence of a single qubit immersed in the XX bath. The decoherence dynamics is found to depend on the internal phases of the XX bath. The short time dynamics of the decoherence factor behaves like a Gaussian, with the decay rate only depending on the filling number of the ground state of the bath. By considering two copies of our qubit-bath systems, we further study the disentanglement dynamics of the two initially entangled qubits coupled to their individual baths. We analytically show that the concurrence is bounded from above by the decoherence factor of a single qubit at all times. The initial Bell state considered suffers from ESD when the XX bath is in its critical phase. We also obtain the disentanglement time in the sudden death region and find that ESD always occurs earlier than the onset of decoherence in a single qubit.
\par The rest of the paper is structured as follows: In Sec. II, we introduce our model Hamiltonian and describe how to obtain the exact qubit-bath time-dependent wavefunctions in the momentum space of the XX spin chain. The components of the Bloch vector of the qubit are obtained by tracing out the bath degrees of freedom over these total wavefunctions. The results for the nonresonant and interacting cases are presented. In Sec. III, we study single qubit decoherence in a single qubit-bath system and disentanglement of two initially entangled qubits in two independent qubit-bath systems. Conclusions are drawn in Sec. IV.

\section{II. Model and Rabi oscillations of a single qubit}
Our model consists of a single qubit coupled to a spin bath of $N$ spins-$1/2$ via the Hamiltonian
\begin{eqnarray}\label{single}
H &=&H_{S}+H_{B}+H_{SB},\nonumber\\
H_S&=&\frac{\omega}{2}(\sigma _z+1),\nonumber\\
H_B&=&\frac{1}{2}\sum^N_{i,j=1}J_{ij} \sigma^+_i\sigma^-_{j}-\frac{h}{2}\sum^N_{j=1}(\sigma^z_{j }+1),\nonumber\\ H_{SB}&=&\sum^N_{j=1}g_j(\sigma^+_{j}\sigma_-+\sigma^-_{j}\sigma_+),
\end{eqnarray}
where $\sigma^{\pm}_j=(\sigma^x_j\pm i\sigma^y_j)/2$ and $\sigma_{\pm}=(\sigma_x\pm i\sigma_y)/2$ are the Pauli matrices for spin $j$ in the spin bath and the central spin, respectively. $\omega$ is the energy difference of the two levels of the single qubit, and $J_{ij}$ is the interaction between bath spins $i$ and $j$. An external magnetic field $h$ in the spin bath is also included. The single qubit is coupled with spin $j$ in the spin bath via XX-type interactions with coupling strength $g_j$. We introduce the collective angular momentum operator $\mathbf{L}=\sum_i\vec{\sigma}_i/2$, where $\vec{\sigma}_i=(\sigma^x_i,\sigma^y_i,\sigma^z_i)$. Note that our $H_{SB}$ takes the same form as that in Ref.~\cite{CR2013} and~\cite{Goan2007}. However, there is no intrabath interaction in Ref.~\cite{CR2013} and uniform intrabath interactions in Ref.~\cite{Goan2007}. In this work, we will choose as the spin bath a periodic one-dimensional chain with nearest neighbor interaction $J_{ij}=J_{ji}=J\delta_{i+1,j}$, namely, an XX spin chain with periodic boundary conditions. For this system, it can be easily checked that the total magnetization $M=\sigma_z/2+L_z$ is a good quantum number. However, the total angular momentum $\mathbf{L}^2=L^2_x+L^2_y+L^2_z$ of the spin bath is not conserved due to either the finite interaction $J$ or the inhomogeneous coupling $g_j$.
\par A spin coherent state of the spin bath, which lives in the $l=N/2$ subspace and is parameterized by the unit vector  $\hat{\Omega}=(\sin\theta\cos\phi,\sin\theta\sin\phi,\cos\theta)$, can be written as~\cite{1972}
\begin{eqnarray}
|\hat{\Omega}\rangle=e^{-iL_z\phi}e^{-iL_y\theta}|\frac{N}{2},\frac{N}{2}\rangle=\sum^N_{n=0}C_n|D^{(\frac{N}{2})}_n\rangle,
\end{eqnarray}
where $C_n=\frac{z^n}{(1+|z|^2)^{N/2}}\sqrt{C^n_N}$ with $z=\cot\frac{\theta}{2}e^{-i\phi}$, and $|D^{(l)}_n\rangle=|l,n-l\rangle$ ($n\in\{0,1,...,2l\}$) are the fully symmetric Dicke states~\cite{1972}, which are simultaneous eigenstates of $\mathbf{L}^2$ and $L_z$ with eigenvalues $l(l+1)$ and $n-l$. To study the Rabi oscillations of the qubit, the initial state is chosen as the product state
\begin{eqnarray}\label{initial}
|\psi(0)\rangle=|1\rangle\otimes|\hat{\Omega}\rangle,
\end{eqnarray}
with the qubit in its up state $|1\rangle$. (The down state will be denoted by $|\bar{1}\rangle$).
\par It will be convenient to work in the interaction picture with respect to $H_{S}+H_{B}$. The energy levels and eigenstates of $H_{B}$ can be obtained by using the Jordan-Wigner transformation $\sigma^-_i=\prod^{i-1}_{j=1}(1-2c^\dag_jc_j)c_i,~\sigma^z_i=2c^\dag_ic_i-1$, where $c_i$ are fermionic operators. $H$ then describes a qubit immersed in a spinless fermion bath,
\begin{eqnarray}\label{1q1f}
H&=&\frac{\omega}{2}(\sigma_z+1)+\frac{J}{2}\sum^N_{j=1}(c^\dag_jc_{j+1}+c^\dag_{j+1}c_j)-h\sum^N_{j=1}c^\dag_j c_j\nonumber\\ &&+\sum^N_{j=1}g_j(c^\dag_jT_j\sigma_-+c_jT^\dag_j\sigma_+),
\end{eqnarray}
with the string operators $
T_j=e^{i\pi\sum^{j-1}_{l=1}c^\dag_lc_l}
$.
One can define two projection operators, $
P_+=\frac{1+T_{N+1}}{2},$ and $P_-=\frac{1-T_{N+1}}{2}
$, which
project onto subspaces where the total fermion number operator $\mathcal{N}_f=\sum^{N}_{l=1}c^\dag_lc_l$ has even or odd eigenvalues $N_f$. For even or odd $N_f$, anti-periodic $c_{N+1}=-c_1$  or periodic  boundary  conditions $c_{N+1}=c_1$, respectively, are imposed on the fermions. As a result, we can introduce the following two sets of Fourier transformations,
\begin{eqnarray}\label{FT}
c_j=\frac{1}{\sqrt{N}}\sum_{k\in K_+}e^{ikj}c_k=\frac{1}{\sqrt{N}}\sum_{k\in K_-}e^{ikj}d_k,
\end{eqnarray}
where $\{c_k\}$ and $\{d_k\}$ are Fourier modes with wave-numbers surviving in $K_+=\{-\pi+\frac{\pi}{N},...,-\frac{\pi}{N},\frac{\pi}{N},...,\pi-\frac{\pi}{N}\}$ and $K_-=\{-\pi,-\pi+\frac{2\pi}{N},...,0,...,\pi-\frac{2\pi}{N}\}$, respectively. Now $H_B$ is diagonalized as
\begin{eqnarray}\label{HXX}
H_{B}&=&P_+H_+P_++P_-H_-P_-,\nonumber\\
H_+&=&\sum_{k\in K_+}\varepsilon_kc^\dag_kc_k,\nonumber\\
H_-&=&\sum_{k\in K_-}\varepsilon_kd^\dag_kd_k,
\end{eqnarray}
with $\varepsilon_k=J\cos k-h$ the single particle spectrum. By direct calculation, we arrive at the interaction picture Hamiltonian
\begin{eqnarray}\label{HI}
&&H_I(t)=e^{i(H_{S}+H_{B})t}H_{SB}e^{-i(H_{S}+H_{B})t}\nonumber\\
&=& \sum^N_{j=1}g_j[(P_-e^{iH_-t}c^\dag_jT_je^{-iH_+t}P_++P_+e^{iH_+t}c^\dag_jT_j e^{-iH_-t}P_-)\nonumber\\
&&\sigma_-e^{-i\omega t}+\rm{H.c.}],
\end{eqnarray}
where H.c. stands for the Hermitian conjugate. Since $M$ is a conserved quantity, it is sufficient to study time evolution from the states $|\psi^{(n)}(0)\rangle=|1\rangle\otimes|D^{(\frac{N}{2})}_n\rangle$. In the interaction picture, the state evolved from the initial state in Eq. (\ref{initial}) then reads $|\psi_I(t)\rangle=\sum_nC_n|\psi^{(n)}(t)\rangle$, with $|\psi^{(n)}(t)\rangle=\mathcal{T}e^{-i\int^t_0ds H_I(s)ds}|\psi^{(n)}(0)\rangle$. In general, the structure of this state is highly complicated due to the non-conservation of the total angular momentum of the spin bath (see Appendix A). To this end, we represent the Dicke states in terms of the momentum space fermion operators:
\begin{eqnarray}\label{Dicke}
|D^{(\frac{N}{2})}_n\rangle&=&\frac{1}{\sqrt{C^n_N}}\sum_{j_1<j_2<...<j_n}\sigma^+_{j_1}...\sigma^{+}_{j_n}|\bar{1}...\bar{1}\rangle\nonumber\\
&=&\frac{1}{\sqrt{C^n_N}}\sum_{j_1<j_2<...<j_n}c^\dag_{j_1}...c^\dag_{j_n}|0\rangle\nonumber\\
&=&\frac{1}{\sqrt{C^n_N}}\sum_{k_1<k_2...<k_n} \sum_{j_1<j_2<...<j_n}\nonumber\\
&&S^*(k_1,...,k_n;j_1,...,j_n) a^\dag_{k_1}...a^\dag_{k_n}|0\rangle,
\end{eqnarray}
where $a_k=c_k(d_k)$ for even (odd) $n$. Here $|0\rangle$ is the vacuum state of the fermions, which corresponds to the state with all bath spins in their down states $|\bar{1}...\bar{1}\rangle$. The function
\begin{eqnarray}
&& S(k_1,...,k_m;j_1,...,j_m)=\nonumber\\
&&\left(\frac{1}{\sqrt{N}}\right)^m\det\left(
                                                  \begin{array}{cccc}
                                                    e^{ik_1j_1} & e^{ik_1j_2} & . & e^{ik_1j_m} \\
                                                    e^{ik_2j_1} & e^{ik_2j_2} & . & e^{ik_2j_m} \\
                                                    . & . & . & . \\
                                                    e^{ik_mj_1} & e^{ik_mj_2} & . & e^{ik_mj_m} \\
                                                  \end{array}
                                                \right),
\end{eqnarray}
is the Slater determinant made up of plane waves.
\par Therefore, $|\psi^{(n)}(t)\rangle$ can be written as a linear combination of free fermion states. In the following we treat even $n$ or odd $n$ separately.\\
\\
(1) $n=even$.\\
It is easily seen that the most general form of $|\psi^{(n)}(t)\rangle$ is
\begin{eqnarray}\label{psieven}
&&|\psi^{(n)}(t)\rangle=|1\rangle\otimes\sum_{k_1<...<k_n}B(k_1,...,k_n;t)\prod^n_{l=1}c^\dag_{k_l}|0\rangle\nonumber\\
&&+|\bar{1}\rangle\otimes\sum_{k_1<...<k_{n+1}}D(k_1,...,k_{n+1};t)\prod^{n+1}_{l=1}d^\dag_{k_l}|0\rangle,
\end{eqnarray}
where $B(k_1,...,k_n;t)$ and $D(k_1,...,k_{n+1};t)$ are coefficients to be determined by the time-dependent Schr{\"o}dinger equation $i\partial_t|\psi^{(n)}(t)\rangle=H_I(t)|\psi^{(n)}(t)\rangle$. After a straightforward calculation (see Appendix B), we arrive at the following two sets of equations of motion for the coefficients $B$ and $D$
\begin{eqnarray}\label{D1}
&&i\dot{D}(p_1,...,p_{n+1};t)= e^{-i\omega t}e^{i\sum^{n+1}_{l=1}\varepsilon_{p_l}t}\sum_{k_1<...<k_{n}}e^{-i\sum^{n}_{l=1}\varepsilon_{k_l}t}\nonumber\\
&&B(k_1,...,k_{n};t)\tilde{f}^*(p_1,...,p_{n+1};k_1,...,k_{n}; \{g_j\} ),\nonumber\\
\end{eqnarray}
\begin{eqnarray}\label{B1}
&&i\dot{B}(p_1,...,p_{n};t)= e^{i\omega t} e^{i\sum^{n}_{l=1}\varepsilon_{p_l}t} \sum_{k_1<...<k_{n+1}}e^{-i\sum^{n+1}_{l=1}\varepsilon_{k_l}t}\nonumber\\
&& D(k_1,...,k_{n+1};t)\tilde{f}(k_1,...,k_{n+1};p_1,...,p_{n}; \{g_j\} ),\nonumber\\
\end{eqnarray}
where the auxiliary function $\tilde{f}$ is defined to be
\begin{eqnarray}\label{ff}
&&\tilde{f}(k_1,...,k_{m+1};p_1,...,p_m;\{g_j\})\nonumber\\
&=&\sum_{j_1<j_2<...<j_{m+1}}   S(k_1,...,k_{m+1};j_1,...,j_{m+1})\nonumber\\
&& \sum^{m+1}_{l=1}g_{j_l}S^*(p_1,...,p_m;j_1,...,\underline{j_l},...,j_{m+1}).
\end{eqnarray}
Here $(j_1,...,\underline{j_l},...,j_{m+1})$ is the string of length $m$, $(j_1,...,j_{l-1},j_{l+1},...,j_{m+1})$, where $j_l$ has been removed. Note that the qubit-bath coupling configuration $\{g_j\}$ is completely incorporated into the the $\tilde{f}$-functions. For simplicity, we take uniform coupling $g=g_j$ in the following numerical calculations. In this case, Eq. (\ref{ff}) can be factorized as $\tilde{f}(k_1,...,k_{m+1};p_1,...,p_m;g)=gf(k_1,...,k_{m+1};p_1,...,p_m)$, where $f$ is the interaction-independent part of the $\tilde{f}$-function.
\par Eqs. (\ref{D1}) and (\ref{B1}) imply that $h$ and $\omega$ only enter the equations of motion through their sum $h+\omega$, the detuning.  The initial values of the $B$s and $D$s can be read off from Eq.~(\ref{Dicke}):
\begin{eqnarray}
B(k_1,...,k_n;0)&=&\frac{1}{\sqrt{C^n_N}}\sum_{j_1<j_2<...<j_n}S^*(k_1,...,k_n;j_1,...,j_n),\nonumber\\
D(k_1,...,k_n;0)&=&0.
\end{eqnarray}
Note that the above equations also include the case of $n=0$, where there are no $k$-arguments for $B(;t)$. The corresponding $f$ function is defined by $f(k;)=\sum_{j}S(k;j)$.\\
\\
(2) $n=odd$.\\
Similarly, the time-evolved states for odd $n$ are of the form
\begin{eqnarray}\label{psiodd}
&&|\psi^{(n)}(t)\rangle=|1\rangle\otimes\sum_{k_1<...<k_n}B'(k_1,...,k_n;t)\prod^n_{l=1}d^\dag_{k_l}|0\rangle\nonumber\\
&&+|\bar{1}\rangle\otimes\sum_{k_1<...<k_{n+1}}D'(k_1,...,k_{n+1};t)\prod^{n+1}_{l=1}c^\dag_{k_l}|0\rangle,
\end{eqnarray}
where the coefficients $B'$ and $D'$ obey the same sets of equations of motion Eqs. (\ref{D1}-\ref{B1}), except the number of arguments for the $B'$s and $D'$s change.
\par
\begin{figure}
  \includegraphics[scale=0.45, angle=0]{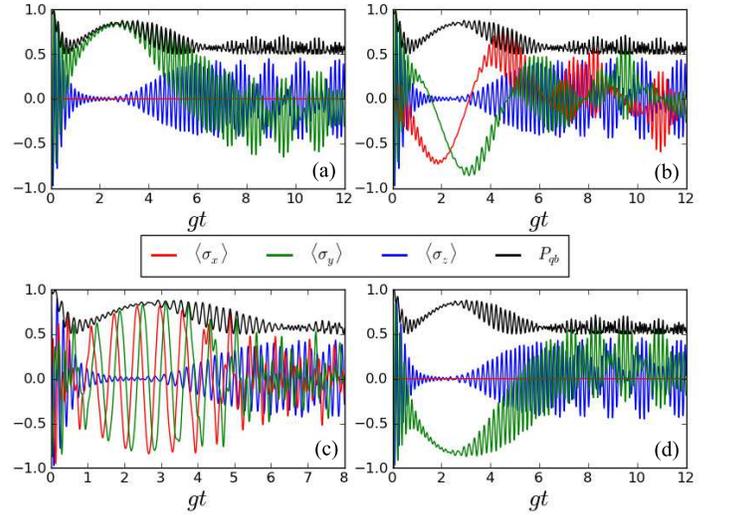}
  \caption{\label{res}Dynamics of $\langle\sigma_x(t)\rangle$ (red), $\langle\sigma_y(t)\rangle$ (green), $\langle\sigma_z(t)\rangle$ (blue) and purity $P_{\rm{qb}}(t)$ (black) in the resonant case $(h+\omega)/g=0$ for four different sets of parameters: (a): $\omega/g=0, z=0.6$, (b): $\omega/g=1, z=0.6$, (c): $\omega/g=10, z=0.6$, (d) $\omega/g=0, z=1.6$. Other parameters: $N=40, J/g=0$. }
\end{figure}

 To get an intuitive understanding of the dynamics, we first consider the non-interacting case $J=0$. In this case the total angular momentum $\mathbf{L}^2$ is conserved and the analytical expression for the Bloch vector $\langle\vec{\sigma}(t)\rangle=\{\langle\sigma_x(t)\rangle,\langle\sigma_y(t)\rangle,\langle\sigma_z(t)\rangle\}$ and the qubit purity $P_{\rm{qb}}(t)=\frac{1}{2}(1+\sum_{i=x,y,z}\langle\sigma_i(t)\rangle^2)$ can be easily calculated (see Appendix A). It turns out that $\langle\sigma_x(t)\rangle$ and $\langle\sigma_x(t)\rangle$ depend on both the detuning $h+\omega$ and the qubit energy difference $\omega$, but $\langle\sigma_z(t)\rangle$ and $P_{\rm{qb}}(t)$ depend only on $h+\omega$. Fig.~1 shows the dynamics of these four quantities in the resonant case $h+\omega=0$, where clear CR behavior appears for the polarization dynamics $\langle\sigma_z(t)\rangle$. This observation has been recently made in Ref. \cite{CR2013} via a correspondence between the non-interacting qubit-spin-bath system and the JC model at large $N$. As in the JC model \cite{CR1990}, in the collapse regime $gt\approx 2.5$ the polarization only undergoes very small oscillations with nearly vanishing amplitudes, but is accompanied by a maximum of the purity. We will refer to such CR behavior as `conventional' CR dynamics observed in the qubit-field system. This behavior can be understood from examining the dynamics of $\langle\sigma_x(t)\rangle$ and $\langle\sigma_y(t)\rangle$. For example, $\langle\sigma_x(t)\rangle$ always vanishes for $\omega=0$ (Fig.~\ref{res}(a)) while $\langle\sigma_y(t)\rangle$ reaches its maximum in the collapse regime, indicating the approximate creation of a pure state $|+\hat{y}\rangle$ with the qubit pointing along the $+\hat{y}$ direction. We observe that $\omega$ controls the frequency of rotation of the Bloch vector in the $x-y$ plane in the conventional collapse region. It was argued in Ref.~\cite{CR2013} that the correspondence between the qubit-spin bath model and the JC model only holds for the parameter regime $|z|^2\ll1\ll N$, and may break down for $|z|^2\ge 1$. In Fig.~\ref{res}(d), we display the dynamics for $z=1.6$ with all the other parameters the same as in Fig.~\ref{res}(a). We see that the CR dynamics still survives and that the behavior of $\langle\sigma_z(t)\rangle$ and purity is almost the same as in Fig.~\ref{res}(a), but with the qubit evolving into state $|-\hat{y}\rangle$ in the collapse regime.
\begin{figure}
  \includegraphics[scale=0.45, angle=0]{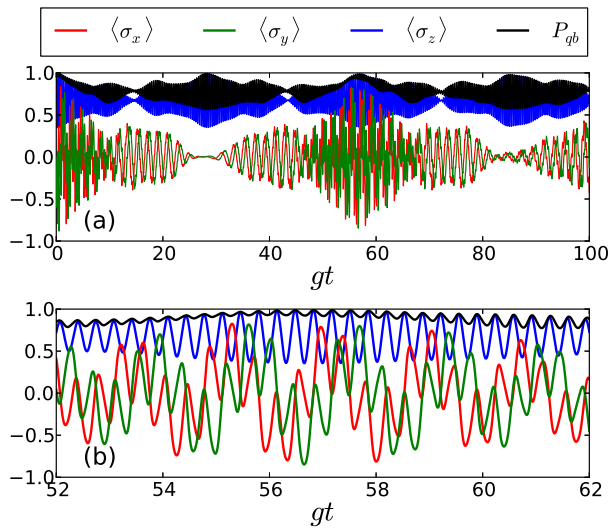}
  \caption{\label{nonres}(a) Dynamics of $\langle\sigma_x(t)\rangle$ (red), $\langle\sigma_y(t)\rangle$ (green), $\langle\sigma_z(t)\rangle$ (blue) and purity $P_{\rm{qb}}(t)$ (black) in the nonresonant case. Parameters: $(h+\omega)/g=15, N=10, J/g=0, z=1,\omega=0$; (b) Magnification of the interval $gt\in[52,62]$, where approximate pure states are sustained during the evolution.}
\end{figure}
\par Fig.~\ref{nonres} shows the results for the nonresonant and noninteracting case $(h+\omega)/g\neq0, J/g=0$. From examining different values of the detuning $h+\omega$, we observe that it controls both the amplitude and period of the oscillations of the envelope of $\langle\sigma_z(t)\rangle$. Larger values of $(h+\omega)/g$ lead to longer periods and smaller amplitudes of these oscillations, as can also be seen from Eq. (\ref{szj0}). Interestingly, the CR dynamics emerges even for a relatively small number of bath spins $N=10$, which would not occur in the resonant case. However, this is not the conventional CR dynamics as seen in the resonant case. For $(h+\omega)/g=15$, there is a collapse region for $\langle\sigma_x\rangle$ and $\langle\sigma_y\rangle$ at $gt\approx30$, where both the purity and $\langle\sigma_z(t)\rangle$ suffer from rapid oscillations between $0.5$ and $1$. More interesting dynamics appears at $gt\approx55 $ (Fig.~\ref{nonres}(b)), where the purity undergoes small oscillations but remains close in absolute value to unity. We will refer to the behavior in both these regions as `unconventional' CR dynamics. Unlike in the resonant case, where the pure state of the qubit rotates in the $x-y$ plane, here the qubit moves along the surface of the northern hemisphere of the Bloch sphere.
\par For finite $J$, although a closed form solution to the equations of motion cannot be obtained, we have been able to solve Eqs. (\ref{D1}-\ref{B1}) numerically for finite $N$. To carry out the integration in a reasonable amount of time, it is necessary to solve for the auxiliary $f$-functions beforehand, and we were able to write a recursive function to do so. This is the most time consuming step in the numerics, and prevented us from examining systems with larger $N$. Based upon the zeros of the $f$-functions, one can also decouple the system of equations in both Eqs. (\ref{D1}-\ref{B1}). Doing so allows each component to be solved in parallel, resulting in a great speedup of the numerical integration.
\par
\begin{figure}
  \includegraphics[width=.5\textwidth]{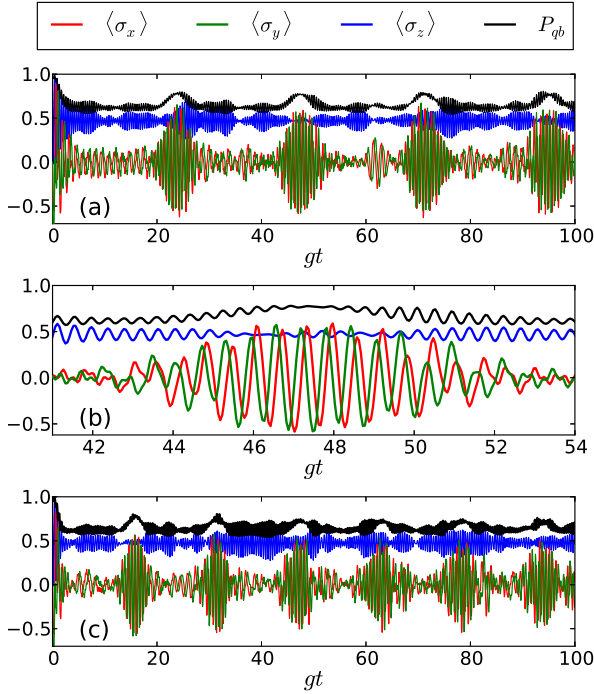}
  \caption{\label{Jhw} The dynamics of the qubit in the nonresonant regime, with $(h+\omega)/g=10$, when the interbath interactions are present. In (a) and (b), $J=0.5$ and in (c), $J=1.0$. (b) displays the conventional collapse region which appears when the interactions are turned on even for $N=10$. Other parameters: $\omega/g=0$, $z=1$.}
\end{figure}

 The three components of the Bloch vector can be calculated from Eq. (\ref{psieven}) and Eq. (\ref{psiodd}), and their expressions in terms of the coefficients $B$, $D$, $B'$ and $D'$ are listed in Appendix A. Numerical results for finite intrabath interaction and finite detuning with $J/g=0.5$ and $(h+\omega)/g=10$ are plotted in Fig.~\ref{Jhw}(a). Comparing with Fig.~2, we see that conventional CR dynamics in $\langle\sigma_z(t)\rangle$ reappears after introducing finite intrabath coupling. This is shown more clearly in Fig.~\ref{Jhw}(b). Except for the facts that the oscillation center of $\langle\sigma_z(t)\rangle$ moves to around $0.5$, and that the peaks of the purity are below 1.0 in this case, the dynamics in the collapse region closely mimics that of in Fig.~1(c). This is an intriguing observation, considering that our spin bath contains only a relatively small number of spins ($N=10$). However, not every peak of the purity is accompanied by the collapse of $\langle\sigma_z(t)\rangle$, as can be seen from the first and third peaks in Fig.~\ref{Jhw}(a). In Fig.~\ref{Jhw}(c), we display the same plot for $J/g=1.0$. We see that increasing the coupling strength $J/g$ causes the period between successive peaks of the purity to decrease. These revivals in the purity also appear to wash out more quickly than they do in Fig.~\ref{Jhw}(a).
\par Finally, we note that for $J=0$ the polarization dynamics $\langle\sigma_z(t)\rangle$ is symmetric under changing the sign of the detuning $h+\omega\to-(h+\omega)$ for fixed $\omega$ (cf. Eq. (\ref{szj0})). However, this is not the case for finite $J$ and only the relative sign between $h+\omega$ and $J$ is relevant (see the end of Appendix A for an example of $N=2$).

\section{III. Entanglement dynamics of two qubits coupled to two individual spin baths}
In the previous section, we studied the reduced dynamics of a single qubit coupled to an interacting spin bath, with the bath initially prepared in the spin coherent state. Now we consider two such copies of the qubit-bath system, between which there is no direct interaction:
 \begin{eqnarray}
H=\sum_{q=1,2} (H^{(q)}_S+H^{(q)}_B+H^{(q)}_{SB}),
\end{eqnarray}
with $H^{(q)}_S, H^{(q)}_B$ and $H^{(q)}_{SB}$ given by Eq. (\ref{single}), and the upper index indicating the operators for copies $q=1$ or $2$.
\par As shown in Ref. \cite{Compagno2007}, the reduced dynamics of the two qubits can be determined completely from that of only one of the two copies.
Explicitly, let $\rho(t)$ denote the reduced density matrix of the two qubits. Assuming a separable initial state $\rho_{\rm{tot}}(0)=\rho(0)\otimes\rho^{(1)}_{B }\rho^{(2)}_{B}$, $\rho(t)$ can be written in the basis of the two qubits $\{|11\rangle,|1\bar{1}\rangle,|\bar{1}1\rangle,|\bar{1}\bar{1}\rangle\}$ as
\begin{eqnarray}\label{aabb}
  \rho_{aa',bb'}(t) =\sum_{cc',dd'}  W^{(1)}_{abcd}(t)W^{(2)}_{a'b'c'd'} (t)\rho_{cc',dd'}(0),
\end{eqnarray}
where
$W^{(q)}_{abcd}(t)$
is determined by the dynamics of each part through
\begin{eqnarray}\label{rhoW}
\rho^{(q)}_{ab}(t)&=&\sum_{cd}W^{(q)}_{abcd}(t) \rho^{(q)}_{cd}(0) ,~q=1,2
\end{eqnarray}
for an initial state $\rho^{(q)} (0)\otimes\rho^{(q)}_B$ of copy $q$.

Thus, in the following we focus on the dynamics of a single qubit coupled to a single bath described by Eq. (\ref{single}), and drop the upper index $q$ for simplicity. Obviously, the dynamics depends on the initial state of the bath $\rho_{B}$. In this section, we will choose the ground state of the isolated XX chain as the bath's initial state. We set $J=-1$ and $h\geq0$ henceforth. For a chain with a finite number of sites, lowering $h$ from the critical field $h_c=1$ to $h=0$ causes $N/2$ level crossings, which correspond to transitions between different parity sectors. This leads to the Kosterlitz-Thouless phase transition in the thermodynamic limit. Correspondingly, the ground states are filled by $N_f=m+1$ $d$-fermions ($c$-fermions) for $m$ even (odd). The detailed ground state structure of the periodic XX chain can be found in Appendix C.

\par In order to get a better understanding of the relationship between decoherence and disentanglement, which is believed to be of importance for both the foundation of quantum mechanics and practical applications of quantum information~\cite{eberly}, we first study the decoherence dynamics of a single qubit.
\subsection{A. Single qubit decoherence}
We suppose that initially the qubit is not entangled with the XX bath. That is,
\begin{eqnarray}\label{phi0}
|\phi(0)\rangle=(a_{\bar{1}} |\bar{1}\rangle+a_1|1\rangle)\otimes |g_{XX}\rangle,
\end{eqnarray}
where $|g_{XX}\rangle$ is the ground state of the XX chain. The coefficients $a_{\bar{1}}$ and $a_1$ satisfy $|a_{\bar{1}}|^2+|a_1|^2=1$. Note that $|g_{XX}\rangle$ is not an eigenstate of Eq. (\ref{single}), so the evolution starting from $|\phi(0)\rangle$ is non-trivial. The spin-flip qubit-bath coupling will induce entanglement between the qubit and spins in the XX chain. The ground state $|g_{XX}\rangle$ characterized by the number of excitations $N_f$ will evolve into superpositions of states within subspaces with $N_f\pm 1$ excitations due to the interaction term $H_{SB}$.
\par Let us first focus on the case of $0<h<1$, where the excitation number $N=m+1\le N-1$. Depending on the parity of the filling number $m$, we will use indices `$o$' or `$e$' to indicate quantities corresponding to odd or even $m$. For $|g_{XX}\rangle=|g_m\rangle_o$ (see Eq. (\ref{gc})) with $m$ odd, the most general form of $|\phi_I(t)\rangle$ will be
\begin{eqnarray}\label{psiIo}
&&|\phi_I(t)\rangle_{o}=\sum_{k_1<...<k_{m+1}}[a_{\bar{1}}A(k_1,...,k_{m+1};t)|\bar{1}\rangle\nonumber\\
&&+a_1B(k_1,...,k_{m+1};t)|1\rangle]\prod^{m+1}_{l=1}c^\dag_{k_l}|0\rangle\nonumber\\
&&+\sum_{k_1<...<k_{m+2}}a_1D(k_1,...,k_{m+2};t)|\bar{1}\rangle\prod^{m+2}_{l=1}d^\dag_{k_l}|0\rangle\nonumber\\
&&+\sum_{k_1<...<k_{m}}a_{\bar{1}}C(k_1,...,k_{m};t)|1\rangle\prod^{m}_{l=1}d^\dag_{k_l}|0\rangle.
\end{eqnarray}
By similar calculations as in the spin coherent state case, we find that $B$ and $D$ obey the same set equations of motion as Eqs. (\ref{D1}) and (\ref{B1}). In addition, the equations of motion for $A$ and $C$ read
\begin{eqnarray}\label{C}
&&i\dot{C}(p_1,...,p_{m};t)=
g e^{i\omega t}e^{i\sum^{m}_{l=1}\varepsilon_{p_l}t}  \sum_{k_1<...<k_{m+1}}\nonumber\\
&& e^{-i\sum^{m+1}_{l=1}\varepsilon_{k_l}t}A(k_1,...,k_{m+1};t)f(k_1,...,k_{m+1};p_1,...,p_m),\nonumber\\
\end{eqnarray}
\begin{eqnarray}\label{A}
&&i\dot{A}(p_1,...,p_{m+1};t)=g e^{-i\omega t}e^{i\sum^{m+1}_{l=1}\varepsilon_{p_l}t}\sum_{k_1<...<k_{m}} \nonumber\\
&&  e^{-i\sum^{m}_{l=1}\varepsilon_{k_l}t}C(k_1,...,k_{m};t)f^*(p_1,...,p_{m+1};k_1,...,k_m).\nonumber\\
\end{eqnarray}
The nonzero initial values of these variables can be read from Eq. (\ref{gd}) and Eq. (\ref{gc})
\begin{eqnarray}\label{IC}
&&A(-m\frac{\pi}{N},...,m\frac{\pi}{N};0)=B(-m\frac{\pi}{N},...,m\frac{\pi}{N};0)=1.\nonumber\\
\end{eqnarray}
All other initial values of $A$, $B$, $C$ and $D$ vanish.
For $h\ge1$, the ground state is the fully polarized state with $m=N-1$, which can be included in Eq. (\ref{psiIo}).
\par The reduced density matrix of the qubit, and hence the $W$ factors in Eq. (\ref{rhoW}), can be obtained by tracing out the bath degrees of freedom. Note that $[P_+H_+P_+,P_-H_-P_-]=0$, so the trace can be taken over $c$-fermions and $d$-fermions independently: $\rho_{o}(t)=tr_{c,d}(|\phi_S(t)\rangle_{o}~_{o}\langle\phi_S(t)|)$ with the Schr{\"o}dinger picture state given by $|\phi_S(t)\rangle_{o}=e^{-i(H_{S}+H_{B})t}|\phi_I(t)\rangle_{o}$. By using Eq. (\ref{rhoW}), we obtain the $W$ factors
\begin{eqnarray}\label{W}
W^{(o)}_{1111}(t)&=&\sum_{k_1<...<k_{m+1}}|B(k_1,...,k_{m+1};t)|^2, \nonumber\\ W^{(o)}_{11\bar{1}\bar{1}}(t)&=&\sum_{k_1<...<k_{m}}|C(k_1,...,k_{m};t)|^2,\nonumber\\
W^{(o)}_{\bar{1}\bar{1}11}(t)&=&\sum_{k_1<...<k_{m}}|D(k_1,...,k_{m+2};t)|^2, \nonumber\\ W^{(o)}_{\bar{1}\bar{1}\bar{1}\bar{1}}(t)&=&\sum_{k_1<...<k_{m+1}}|A(k_1,...,k_{m+1};t)|^2,\nonumber\\
W^{(o)}_{1\bar{1}1\bar{1}}(t)&=&e^{-i\omega t}\sum_{k_1<...<k_{m+1}}\nonumber\\
&&A^*(k_1,...,k_{m+1};t)B(k_1,...,k_{m+1};t),\nonumber\\
W^{(o)}_{\bar{1}1\bar{1}1}(t)&=& W^{(o)*}_{1\bar{1}1\bar{1}}(t),
\end{eqnarray}
with all other elements vanishing. A similar analysis can be carried out for $m$ even where $|g_{XX}\rangle=|g_m\rangle_e$ (see Appendix C).
\par We recognize the decoherence factor of a single qubit~\cite{Zurek2005} $r(t)$ from Eq. (\ref{rhoW}) as
\begin{eqnarray}\label{rt}
r(t)=W^{(o)}_{1\bar{1}1\bar{1}}(t),
\end{eqnarray}
whose absolute value is bounded by $0\leq|r(t)|^2\leq1$, corresponding to complete decoherence and no loss of coherence, respectively.
\par
\begin{figure}
\includegraphics[width=.5\textwidth]{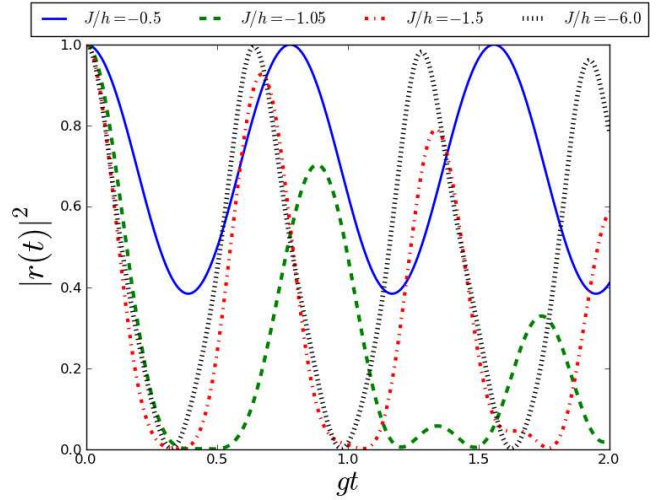}
\caption{\label{weak} The decoherence factor is shown for different values of nearest neighbor coupling $J/h$ in the weak coupling regime with $h/g=10$. Other parameters are: $N=10, \omega/g=0$.}
\end{figure}

 In Fig.~\ref{weak}, we plot the temporal evolution of the decoherence factor $|r(t)|^2$ for different values of nearest neighbor couplings $J/g$ in the weak qubit-bath coupling regime $|J/g|,h/g\gg1$. The revival of coherence occurs since the bath is finite. The loss of coherence is modest for the smallest value of $|J/h|=0.5$, and $|r(t)|^2$ approaches zero only for values $J/h<-1$, namely, in the critical regime of the XX chain.  As the intrabath interaction strength $|J/h|$ is increased, the coherence of the qubit is first suppressed, as can be seen by comparing the curves for $J/h=-1.05$ and $J/h=-0.5$, and then enhanced, as can be seen by examining the curves for $J/h=-1.5$ and $J/h=-6.0$. These observations indicate that the relationship between decoherence and interaction strength is not straightforward.
\par
\begin{figure}
\includegraphics[width=.5\textwidth]{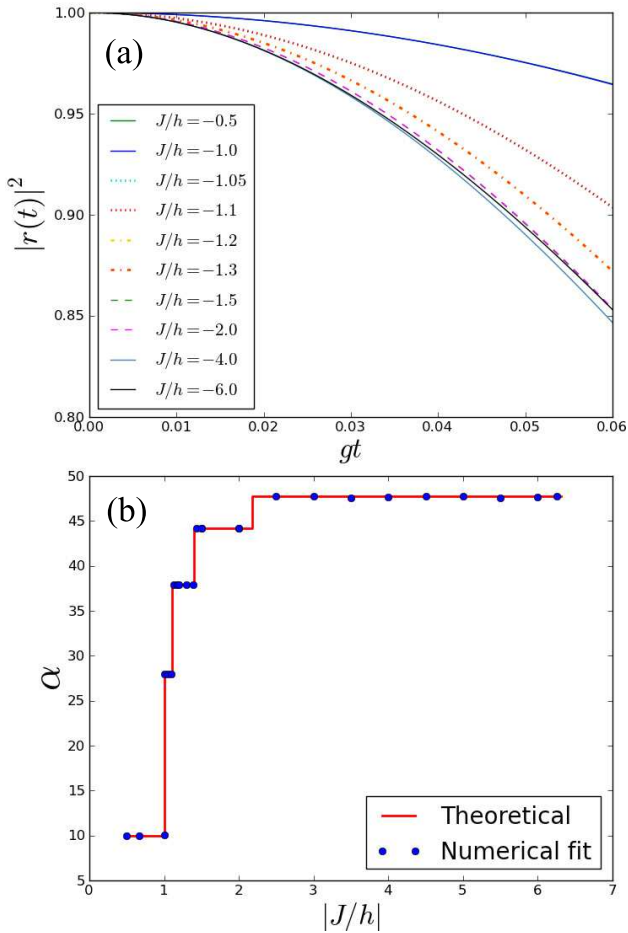}
\caption{\label{gtalpha} (a) The short time behavior of the decoherence factor $|r(t)|^2$. For $gt\ll 1$, the curves collapse onto five Gaussians with decay rates that depend only the value of $m$, the filling number. Two curves within each group of two lines have been plotted with their own line style, and for $gt<0.03$, the five groups can be made out. (b) The dependence of the decay rate $\alpha$ on the intrabath interaction strength $|J/h|$. The red line is the theoretical calculation of Eq. (\ref{alpha}), and the blue dots are numerical fits at small times to the decoherence factor for different values of $J/h$. Other parameters are: $N=10, \omega/g=0$.}
\end{figure}

 Fig.~\ref{gtalpha}(a) displays the short time behavior of the decoherence factor $|r(t)|^2$. It can be seen that when $gt$ is small, $|r(t)|^2$ decays as a Gaussian
\begin{eqnarray}\label{Gass}
|r(t)|^2\sim e^{-\alpha (gt)^2}.
\end{eqnarray}
In Fig.~\ref{gtalpha}(b), we display several values of the exponent $\alpha$ for different values of $|J/h|$ as blue dots which were numerically fit to $|r(t)|^2$ for small times. Interestingly, we note that $\alpha$ exhibits plateaus as a function of $|J/h|$. This behavior can be understood from second-order time-dependent perturbation theory in the qubit-bath coupling $g/J$. It turns out that the initial Gaussian rate is given by (see Appendix D for the derivation)
\begin{figure}
\includegraphics[width=.5\textwidth]{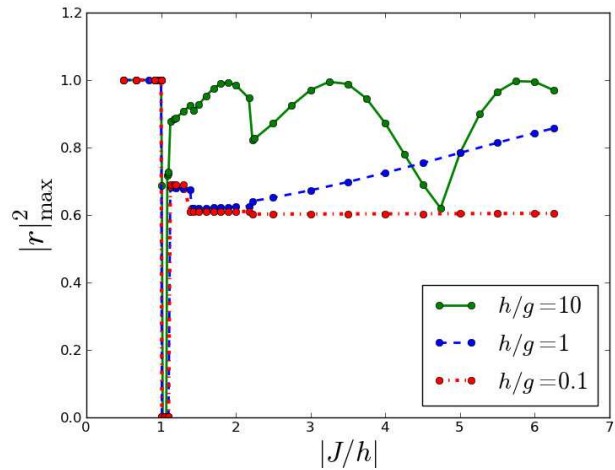}
  \caption{\label{rmax} The first maximum the decoherence factor reaches, $|r|^2_{\max}$, is plotted against the value of $|J/h|$ for different values of $h/g$ corresponding to different qubit-bath couplings. Other parameters are: $N=10, \omega/g=0$.}
\end{figure}
\begin{eqnarray}\label{alpha}
\alpha &=& \sum_{p_1<...<p_m}  |f(k_1,...,k_{m+1};p_1,..,p_m)|^2+\nonumber\\
&&\sum_{p_1<...<p_{m+2}}|f(p_1,...,p_{m+2};k_1,...,k_{m+1})|^2,
\end{eqnarray}
where $(k_1,...,k_{m+1})=(-m\frac{\pi}{N},...,m\frac{\pi}{N})$ or $(-m\frac{\pi}{N},...,0,...,m\frac{\pi}{N})$ for initial states with $|g_{XX}\rangle=|g_m\rangle_o$ or $|g_m\rangle_e$. This perturbative result is displayed as the red set of plateaus in Fig.~\ref{gtalpha}(b). Note that the $f$-functions, and hence the rate $\alpha$, have nothing to do with the system's parameters and only depend on the filling number $m$, which explains the presence of plateaus. In Fig.~\ref{gtalpha}(a), two curves are plotted for each value of $m$, and for $gt<0.03$ the ten curves are seen to collapse into five groups corresponding to the five different values of $m$. The first divergence within a group can be seen for the $m=5$ sector, where the curves for $J/h=-4.0$ and $J/h=-6.0$ separate past $gt\approx 0.03$.

\par The short time behavior of the decoherence factor for intermediate qubit-bath coupling, with $h/g=1$, and strong qubit-bath coupling, with $h/g=0.1$ (not shown here), is similar to that of weak qubit-bath coupling. In particular, they are also characterized by Gaussian behavior. However, the behavior at longer times, as $|J/h|$ is increased, changes. In order to quantitatively compare the behavior of $|r(t)|^2$ in these three regimes, in Fig.~\ref{rmax} we plot the value of the first maximum of the decoherence factor $|r|^2_{\max}$ (aside from the initial value $|r(0)|^2=1$) as a function of the intrabath interaction strength $|J/h|$ for the three coupling regimes examined above. This quantity is representative of the extent to which coherence is  maintained in the qubit~\cite{Relano2008}. For all three regimes, when the bath is in a polarized phase $|J/h|<1$, the decoherence factor returns to unity after one oscillation, and indeed it appears that the periodic revival of the coherence continues for all times. All of the curves exhibit a sudden drop at $|J/h|=1$, reflecting the transition to the critical phase of the bath. Interestingly, in the critical region $|J/h|>1$, $|r|^2_{\max}$ displays markedly different behavior in each regime. For weak qubit-bath coupling, $|r|^2_{\max}$ behaves non-monotonically and oscillates about high values. However, $|r|^2_{\max}$ appears to monotonically increase for $h/g=1$ and $h/g=0.1$, albeit very slowly for the latter. These results seem to show that strong intrabath interaction strength suppresses the decoherence of the qubit, a result has been observed for a qubit coupled to a spin bath with homogeneous self-interaction \cite{JPA2002,Milburn2005}.

\par
\begin{figure}
 \includegraphics[width=.5\textwidth]{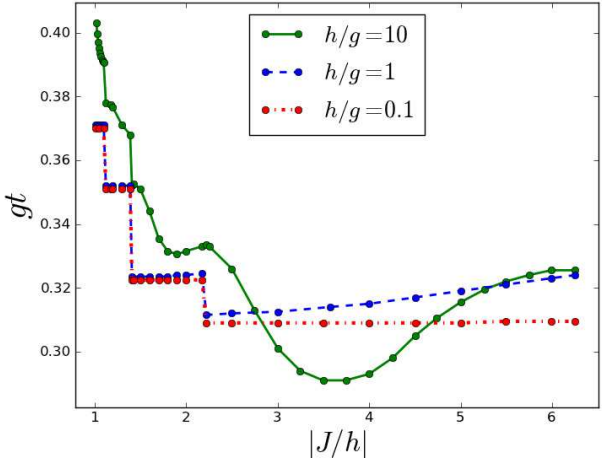}
  \caption{\label{decmin} The time $gt$ at which the decoherence factor $|r(t)|^2$ reaches its first minimum is plotted against the intrabath interaction strength $|J/h|$ for the three qubit-bath coupling regimes. Other parameters are: $N=10, \omega/g=0$.}
\end{figure}

 As another means of assessing the effect of intrabath interactions on the coherence of the qubit, in Fig.~\ref{decmin} we plot the time $gt$ at which the decoherence factor $|r(t)|^2$ reaches its first minimum as a function of $|J/h|$ for the three qubit-bath coupling regimes. We focus on the critical region $|J/h|>1$, as the time of the first minimum for $|J/h|\leq 1$ is much larger and does not display much variation. Interestingly, the green curve, for which $h/g=10$, initially decreases sharply with each successive sector and then displays a global minimum at $|J/h|\approx 3.8$, where the decoherence disappears the quickest. For intermediate and strong qubit-bath coupling $h/g=1$ and $h/g=0.1$, as $|J/h|$ is increased and successive magnetization sectors of the bath are encountered, the time $gt$ of the first minimum drops. But unlike the weak coupling case, $gt$ increases monotonically within each sector, agreeing with previous results that strong interactions within the bath suppress decoherence of the qubit~\cite{JPA2002}.

\subsection{B. Disentanglement of two initially entangled qubits}\label{entanglement}
Now, we turn to the study of the disentanglement of two qubits interacting with independent XX-baths. We focus on one type of initial state for the two-qubit system $|\Psi\rangle=\alpha|\bar{1}1\rangle+\beta|1\bar{1}\rangle$ with $\alpha$ real and $\alpha^2+|\beta|^2=1$. From Eq. (\ref{aabb}), it follows that the time evolved reduced density matrix for the two qubits reads
\begin{eqnarray}
\rho (t)= \left(
          \begin{array}{cccc}
          \rho _{11,11}(t) & 0  & 0  & 0  \\
            0  &  \rho _{1\bar{1},1\bar{1}}(t) & \rho _{1\bar{1},\bar{1}1}(t)  & 0  \\
            0  & \rho^{ *}_{1\bar{1},\bar{1}1}(t)  & \rho _{\bar{1}1,\bar{1}1}(t)  & 0  \\
           0   &0   & 0  &  \rho _{\bar{1}\bar{1},\bar{1}\bar{1}}(t) \\
          \end{array}
        \right),
\end{eqnarray}
with
\begin{eqnarray}
\rho _{11,11}(t)&=&W_{11\bar{1}\bar{1}}(t)W_{1111}(t),\nonumber\\
\rho _{1\bar{1},1\bar{1}}(t)&=&\alpha^2W_{11\bar{1}\bar{1}}(t) W_{\bar{1}\bar{1}11} (t) +|\beta|^2W_{1111} (t) W_{\bar{1}\bar{1}\bar{1}\bar{1}}(t),\nonumber\\
\rho _{1\bar{1},\bar{1}1}(t)&=&\alpha\beta W_{1\bar{1}1\bar{1}}(t) W_{\bar{1}1\bar{1}1} (t) ,\nonumber\\
\rho _{\bar{1}1,\bar{1}1}(t)&=&\alpha^2W_{\bar{1}\bar{1}\bar{1}\bar{1}}(t) W_{1111}(t)  +|\beta|^2W_{\bar{1}\bar{1}11} (t) W_{11\bar{1}\bar{1}}(t),\nonumber\\
\rho _{\bar{1}\bar{1},\bar{1}\bar{1}}(t)&=& W_{\bar{1}\bar{1}\bar{1}\bar{1}}(t)W_{\bar{1}\bar{1}11}(t),
\end{eqnarray}
where we have assumed that the two environments are identical, so that $W_{abcd}(t)=W^{(1)}_{abcd}(t)=W^{(2)}_{abcd}(t)$. We use the concurrence~\cite{wrooter} to measure the bipartite entanglement between the two qubits. The concurrence is defined as
\begin{eqnarray}
C(t)&=&\max\{0,2\lambda_{\max}(t)-tr\sqrt{\rho(t)\tilde{\rho}(t)}\},\nonumber\\
\tilde{\rho}(t)&=&\sigma_y\otimes\sigma_y\rho^*(t)\sigma_y\otimes\sigma_y,
\end{eqnarray}
where $\lambda_{\max}$ is the largest eigenvalue of the matrix $\sqrt{\rho(t)\tilde{\rho}(t)}$. The concurrence for state $\rho(t)$ reads
\begin{eqnarray}\label{Ct}
C(t)&=&\max\{0,2|\alpha\beta||r(t)|^2-2\sqrt{\rho _{11,11}(t)\rho _{\bar{1}\bar{1},\bar{1}\bar{1}}(t)}\},\nonumber\\
\end{eqnarray}
where $|r(t)|^2$ is the single qubit decoherence factor in Eq. (\ref{rt}).
\begin{figure}
\includegraphics[width=.5\textwidth]{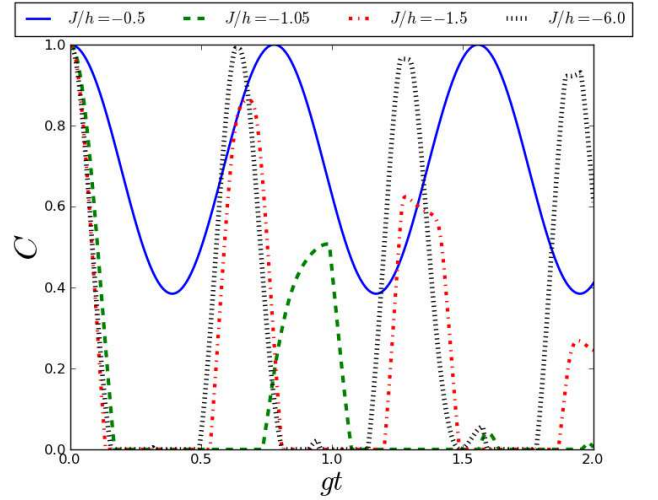}
  \caption{\label{conc} Concurrence dynamics for the initial state $|\Psi\rangle=(|1\bar{1}\rangle+|\bar{1}1\rangle)/\sqrt{2}$ for different values of nearest neighbor coupling $J/h$. The corresponding decoherence factors, plotted in Fig~\ref{weak}, are an upper bound for these curves. When entanglement sudden death is present, it always occurs before the qubits' individual decoherence factors reach a minimum. Other parameters are: $N=10, h/g=10, \omega/g=0$.}
\end{figure}
\begin{figure}
\includegraphics[width=.5\textwidth]{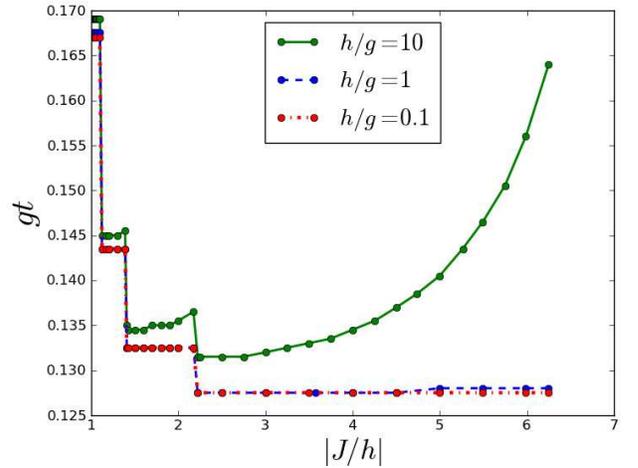}
  \caption{\label{esd} The time $gt$ at which the concurrence first vanishes is plotted against intrabath interaction strength $|J/h|$. Other parameters are: $N=10, \omega/g=0$.}
\end{figure}
\par In Fig.~\ref{conc}, we plot the evolution of concurrence as a function of $gt$ from the maximally entangled Bell state $(|1\bar{1}\rangle+|\bar{1}1\rangle)/\sqrt{2}$. In order to make a comparison between the disentanglement dynamics and the decoherence of a single qubit, we set all parameters to be the same as those in Fig.~\ref{weak}. The concurrence appears to be bounded from above by the corresponding decoherence factor $|r(t)|^2$ for all time. For $0>J/h>-1$, the XX chain is in the fully polarized state along the $+\hat{z}$ direction. The concurrence shows regular oscillations about a high value and never vanishes in this regime. In fact, we have seen that the two-qubit concurrence $C(t)$ exactly coincides with the single qubit decoherence factor $|r(t)|^2$ here. This relation can be understood by examining Eq. (\ref{Ct}): for $2|\alpha\beta||r(t)|^2-2\sqrt{\rho _{11,11}(t)\rho _{\bar{1}\bar{1},\bar{1}\bar{1}}(t)}\leq0$, we have $C(t)=0\leq |r(t)|^2$; while for $2|\alpha\beta||r(t)|^2-2\sqrt{\rho _{11,11}(t)\rho _{\bar{1}\bar{1},\bar{1}\bar{1}}(t)}>0$, we have
\begin{eqnarray}
C(t)&=&2|\alpha\beta||r(t)|^2-2\sqrt{\rho _{11,11}(t)\rho _{\bar{1}\bar{1},\bar{1}\bar{1}}(t)}\nonumber\\
&\leq&2|\alpha\beta||r(t)|^2\leq|r(t)|^2.
\end{eqnarray}
Hence, the concurrence is always bounded from above by $|r(t)|^2$. When the spin bath is in a polarized state, we always have $D(\{k_i\};t)=0$, as can be seen from Eq. (\ref{psiIo}). Therefore, $C(t)=2|\alpha\beta||r(t)|^2=|r(t)|^2$ for the Bell state with $\alpha=\beta=1/\sqrt{2}$. A similar conclusion also holds for the other type of entangled state $\alpha|\bar{1}\bar{1}\rangle+\beta|11\rangle$. This relationship between decoherence and disentanglement has also been observed before in spin-boson type models when the two qubits are coupled to separate bosonic baths~\cite{Tolkunov2005, Ann2007}.
\par On the other hand, ESD always exists in the critical regime $J/h<-1$. This is consistent with the result for two distant qubits coupled locally to an XXZ spin chain via isotropic Heisenberg qubit-bath coupling~\cite{Chen2008}, where it was found that ESD is absent in the ferromagnetic or polarized phase of the spin bath. We also observe that ESD always occurs earlier than the minimum of the corresponding single-qubit decoherence factor, a result in agreement with the case of qubits coupled to independent bosonic baths~\cite{eberly}. In the sudden death region, revival of the entanglement appears a period of time after disentanglement, which is also observed in Ref.~\cite{Chen2008}. This revival phenomenon is induced by the non-Markovian nature of the spin bath~\cite{Petru2004}.
\par In order to compare the entanglement dynamics with the decoherence of a single qubit, we plot the disentanglement time, which is defined as the time when the concurrence first vanishes, as a function of intrabath coupling $|J/h|$ in the sudden death region $|J/h|>1$ in Fig.~\ref{esd}. The time $gt$ until ESD occurs decreases as sectors with lower filling factors $m$ are encountered, but within each sector, $gt$ increases as $J/h$ is increased, although this effect is only pronounced for weak qubit-bath coupling $h/g=10$. This plot displays many similarities to Fig.~\ref{decmin}. However, for the green curve with $h/g=10$, increasing $|J/h|$ only causes an increase in $gt$ with each sector, in contrast to the nonmonotonic behavior of the corresponding curve in Fig.~\ref{decmin}. Also, the intrabath coupling strengths at which the entanglement dies the fastest and at which the decoherence is minimized most quickly are not the same. In spite of these differences, there is a qualitative agreement between Figs.~\ref{decmin} and \ref{esd}, indicating that the decoherence dynamics of a single qubit and the entanglement dynamics of two noninteracting qubits are linked, especially when the qubits are strongly coupled to their respective baths.

\section{IV. Discussion and outlook}
In this work, we studied the reduced dynamics of a specific qubit-spin-bath model. Unlike the spin conserving Ising-type qubit-bath coupling utilized in most previous works, we considered an XX-type spin-flip qubit-bath coupling, which complicates the analytical analysis, since the system-bath interaction term does not commute with the rest of the Hamiltonian. In addition, we model interactions in the bath by introducing nearest-neighbor XX-type couplings among the bath spins. Such a model may be more physical than the non-interacting `spin star' and homogeneously interacting spin baths which have been examined before, but is more difficult to treat analytically. However, by mapping this XX chain into momentum space via the Jordan-Wigner transformation, we have shown how to obtain the equations of motion for the time-dependent total wavefunctions in momentum space. The reduced dynamics of a single qubit is then obtained by tracing out the bath degrees of freedom.
\par Using the above results, we first studied the Rabi oscillations of the qubit with the bath initially prepared in a spin coherent state. Interestingly, the interplay between off-resonance and intrabath interactions was found to produce conventional collapse and revival behavior even for a relatively small spin bath size. We further discussed the bath-induced decoherence of a single qubit with the bath's initial state taken to be its ground state. We found that the decoherence properties of the qubit depend on the internal phases of the XX bath. Specifically, the short time decay rate of the decoherence factor only depends on the filling number of the bath ground state. This result was confirmed through second order time-dependent perturbation theory. Finally, we considered two independent copies of such qubit-bath subsystems and studied the disentanglement dynamics of two initially entangled qubits. The two qubits are always entangled if the XX bath is in its polarized state, whereas entanglement sudden death appears in the critical phase. Qualitative similarities were observed between the time dependence of the two-qubit entanglement and the single-qubit decoherence factor, and we showed that the concurrence is bounded from above by the decoherence factor of the single qubit. However, for spin-boson type systems, it has been shown that this is not the case when the two-qubits share a common bosonic bath~\cite{PRA2009,PRL2003}. We believe that such a relation between single-qubit decoherence and two-qubit disentanglement might also break down for two qubits coupled to a common XX bath. This deserves further study based on our model system.
\par The XX spin chain in our model is equivalent to the one-dimensional Bose-Hubbard model in the hard-core limit~\cite{BH2004}, which can be realized using a cold atomic gas contained in an optical lattice~\cite{Zoller1998,Bloch2003}. Correspondingly, the qubit-spin-bath coupling $H_{SB}$ can be mapped to a conventional spin-boson coupling within the rotating-wave approximation, $\sum_jg_j(b^\dag_j\sigma_-+b_j\sigma_+)$, where $b_j$ are bosonic operators. Regarding the methodology used in this work, we note that the only requirement for our formalism is the conservation of the total magnetization $M$. Thus, our method can also be applied to the nonuniform Heisenberg type qubit-spin-bath coupling~\cite{Sarma2012,FaribaultPRB2013}, where the total magnetization is conserved.
\par Our results suggest that turning on interactions among bath spins can have markedly different effects on the decoherence and entanglement properties of the central spins, depending on how strongly they are coupled to their environment. Therefore, this work may be of relevance to all efforts aimed at using such systems to construct quantum information processing devices.
\\
\\
\noindent{\bf Acknowledgements:}
We acknowledge support from NSF Grant No. CHE-1058644 and ARO-MURI Grant No. W911NF-08-1-0124. A.N. was supported by the Program in Plasma Science and Technology at Princeton University. The calculations in this work were performed at the TIGRESS high performance computer center at Princeton.

\section{Appendix A: Structure of $|\psi^{(n)}(t)\rangle$ and expressions for the Bloch vector $\langle\vec{\sigma}\rangle$}
\par In the non-interacting case $J=0$ with uniform coupling $g_j=g$, the total angular momentum of the spin bath is conserved, so $|\psi^{(n)}(t)\rangle=\mathcal{T} e^{-i\int^t_0dsH_I(s)ds}|1\rangle\otimes|D^{(\frac{N}{2})}_n\rangle$ takes the simple form
\begin{eqnarray}\label{J0}
 |\psi^{(n)}(t)\rangle=a_n(t)|1\rangle|D^{(\frac{N}{2})}_n\rangle+b_n(t)|\bar{1}\rangle|D^{(\frac{N}{2})}_{n+1}\rangle.
\end{eqnarray}
 By applying the Schr{\"o}dinger operator Eq. (\ref{HI}) to the above equation, we obtain the following equations of motion for the coefficients $a_n(t)$ and $b_n(t)$
\begin{eqnarray}
i\dot{a}_n(t)&=&\tilde{g}_ne^{ i(h+\omega)t}b_n(t),\nonumber\\
i\dot{b}_n(t)&=&\tilde{g}_ne^{ -i(h+\omega)t}a_n(t)
\end{eqnarray}
with initial conditions $a_n(0)=1$, $b_n(0)=0$ and $\tilde{g}_n=g\sqrt{(n+1)(N-n)}$. The solutions are
\begin{eqnarray}
a_n(t)&=&e^{\frac{i}{2}(h+\omega)t} [- i(h+\omega)\frac{\sin\frac{t}{2}\sqrt{4\tilde{g}_n^2+(h+\omega)^2}  }{\sqrt{4\tilde{g}_n^2+(h+\omega)^2}}\nonumber\\
&&+ \cos\frac{ t}{2}\sqrt{4\tilde{g}_n^2+(h+\omega)^2}],\nonumber\\
b_n(t)&=&-2i\tilde{g}_ne^{-\frac{i}{2}(h+\omega)t}\frac{\sin \frac{1}{2}\sqrt{4\tilde{g}_n^2+(h+\omega)^2}t}{\sqrt{4\tilde{g}_n^2+(h+\omega)^2}}.
\end{eqnarray}
The polarization dynamics is given by
\begin{eqnarray}\label{szj0}
\langle\sigma_z(t)\rangle&=&\sum^N_{n=0}|C_n|^2[|a_n(t)|^2-|b_n(t)|^2]\nonumber\\
&=&1-8\sum^N_{n=0} \frac{\tilde{g}^2_n|C_n|^2\sin^2\frac{t}{2}\sqrt{4\tilde{g}^2_n+(h+\omega)^2}}{4\tilde{g}^2_n+(h+\omega)^2} .\nonumber\\
\end{eqnarray}
The other two components can be calculated directly from Eq. (\ref{J0})
\begin{eqnarray}
\langle\sigma_x(t)\rangle&=&2\Re[e^{-i\omega t}\sum^N_{n=1}C^*_{n-1}C_nb^*_{n-1}a_n],\nonumber\\
\langle\sigma_y(t)\rangle&=&-2\Im[e^{-i\omega t}\sum^N_{n=1}C^*_{n-1}C_nb^*_{n-1}a_n].
\end{eqnarray}
We also monitor the purity dynamics of the qubit
 \begin{eqnarray}
P_{\rm{qb}}(t)&=&\frac{1}{2}(1+\sum_{i=x,y,z}\langle\sigma_i(t)\rangle^2)\nonumber\\
&=&\frac{1+ \langle\sigma_z(t)\rangle^2}{2}+2|\sum^N_{n=1}C^*_{n-1}C_nb^*_{n-1}a_n|^2.
\end{eqnarray}
Although $\langle\sigma_x(t)\rangle$ and $\langle\sigma_y(t)\rangle$ depend on both $h+\omega$ and $\omega$, $\langle\sigma_z(t)\rangle$ and $P_{\rm{qb}}(t)$ depend only on $h+\omega$. Also, note that $\langle\sigma_z(t)\rangle$ is symmetric under changing the sign of the detuning: $h+\omega\to-(h+\omega)$.
\par For finite $J$ and/or non-uniform coupling $g_j$, the total angular momentum $\mathbf{L}$ of the spin bath is not conserved. So $|\psi^{(n)}(t)\rangle$ will be driven into other $l$-subspaces under the action of $H_I(t)$:
\begin{eqnarray}\label{JF}
  |\psi^{(n)}(t)\rangle&=& |1\rangle\sum^{\frac{N}{2}}_{m=|n-\frac{N}{2}|}a^{(m)}_n(t)|D^{(m)}_n\rangle+\nonumber\\
  &&|\bar{1}\rangle\sum^{\frac{N}{2}}_{m=|n+1-\frac{N}{2}|}b^{(m)}_n(t)|D^{(m)}_{n+1}\rangle,
\end{eqnarray}
which complicates the analysis. In this case, the dynamics of the Bloch vector can be obtained from Eq. (\ref{psieven}) and Eq. (\ref{psiodd}) as
 \begin{eqnarray}
&&\langle\sigma_x(t)\rangle=2\Re[e^{-i\omega t}Z(t)],\nonumber\\
&&\langle\sigma_y(t)\rangle=-2\Im[e^{-i\omega t}Z(t)],
\end{eqnarray}
with
 \begin{eqnarray}
 Z(t)&=&  \sum^{\frac{N}{2}}_{n=1}C^*_{2n-1}C_{2n}\nonumber\\
&&\sum_{k_1<...<k_{2n}}D'^*(k_1,...,k_{2n};t)B(k_1,...,k_{2n};t)+\nonumber\\
&&\sum^{\frac{N}{2}-1}_{n=0}C^*_{2n}C_{2n+1}\nonumber\\
&&\sum_{k_1<...<k_{2n+1}}D^*(k_1,...,k_{2n+1};t)B'(k_1,...,k_{2n+1};t),\nonumber\\
\end{eqnarray}
and
 \begin{eqnarray}
&&\langle\sigma_z(t)\rangle=\langle\psi_I(t)|\sigma_z|\psi_I(t)\rangle\nonumber\\
&=&\sum^{\frac{N}{2}}_{n=0}|C_{2n}|^2[\sum_{k_1<...<k_{2n}}|B(k_1,...,k_{2n};t)|^2\nonumber\\
&&-\sum_{k_1<...<k_{2n+1}}|D(k_1,...,k_{2n+1};t)|^2]\nonumber\\
&+&\sum^{\frac{N}{2}-1}_{n=0}|C_{2n+1}|^2[\sum_{k_1<...<k_{2n+1}}|B'(k_1,...,k_{2n+1};t)|^2\nonumber\\
&&-\sum_{k_1<...<k_{2n+2}}|D'(k_1,...,k_{2n+2};t)|^2].
\end{eqnarray}
Unlike the non-interacting and uniform coupling case, the dynamics of $\langle\sigma_z(t)\rangle$ is no longer symmetric under $h+\omega\to-(h+\omega)$, as can be seen from the a simple example of $N=2$; i.e., for a spin bath made up of only two spins, where an analytical expression can be obtained (with equal qubit-bath coupling $g$):
 \begin{eqnarray}
 \langle\sigma_z(t)\rangle&=&|C_0|^2\frac{8g^2\cos t\sqrt{8g^2+J_-^2}+J_-^2}{8g^2+J_-^2}\nonumber\\
&&+|C_1|^2\frac{8g^2\cos t\sqrt{8g^2+J_+^2}+J_+^2}{8g^2+J_+^2}+|C_2|^2.\nonumber
\end{eqnarray}
with $J_\pm=h+\omega\pm J$. Note that only the relative sign between $h+\omega$ and $J$ is relevant.
\section{Appendix B: Derivation of Eqs. (\ref{D1}-\ref{B1})}
Eqs. (\ref{D1}-\ref{B1}) can be derived by inserting Eq. (\ref{psieven}) into the time-dependent Schr{\"o}dinger equation
\begin{eqnarray}\label{Sch}
i\partial_t|\psi^{(n)}(t)\rangle=H_I(t)|\psi^{(n)}(t)\rangle.
\end{eqnarray}
After acting with $H_I(t)$ on $|\psi^{(n)}(t)\rangle$, only two terms survive:
\begin{eqnarray}
&&H_I(t)|\psi^{(n)}(t)\rangle=|\psi^{(n)}_c(t)\rangle+|\psi^{(n)}_d(t)\rangle, \nonumber\\
&&|\psi^{(n)}_c(t)\rangle= \sum^N_{j=1}g_j P_-e^{iH_-t}c^\dag_jT_je^{-iH_+t}P_+\sigma_-e^{-i\omega t}|1\rangle\nonumber\\
&&\sum_{k_1<...<k_n}B(k_1,...,k_n;t)\prod^n_{l=1}c^\dag_{k_l}|0\rangle,\nonumber\\
&&|\psi^{(n)}_d(t)\rangle= \sum^N_{j=1}g_jP_+e^{iH_+t}c_jT_j e^{-iH_-t}P_-\sigma_+e^{i\omega t}|\bar{1}\rangle\nonumber\\
&&\sum_{k_1<...<k_{n+1}}D(k_1,...,k_{n+1};t)\prod^{n+1}_{l=1}d^\dag_{k_l}|0\rangle.
\end{eqnarray}
$|\psi^{(n)}_c(t)\rangle$ can be calculated as
\begin{eqnarray}
|\psi^{(n)}_c(t)\rangle&=& e^{-i\omega t}|\bar{1}\rangle e^{iH_-t} \sum_{k_1<...<k_n}e^{-i\sum^n_{l=1}\varepsilon_{kl}t} \nonumber\\
&&B(k_1,...,k_n;t)|\chi_{k_1,...,k_n}\rangle,
\end{eqnarray}
with
\begin{eqnarray}
&&|\chi_{k_1,...,k_n}\rangle=\sum^N_{j=1}g_jc^\dag_jT_j\prod^n_{l=1}c^\dag_{k_l}|0\rangle\nonumber\\
&=&\sum^N_{j=1}g_jT_jc^\dag_j\sum_{j_1<j_2...<j_n}S(k_1,...,k_n;j_1,...,j_n)c^\dag_{j_1}...c^\dag_{j_n}|0\rangle\nonumber\\
&=& \sum_{j<j_1<j_2...<j_n}g_j S(k_1,...,k_n;j_1,...,j_n)c^\dag_jc^\dag_{j_1}...c^\dag_{j_n}|0\rangle+...\nonumber\\
&&+\sum_{j_1<...<j_l<j<j_{l+1}...<j_n}g_jS(k_1,...,k_n;j_1,...,j_n) \nonumber\\
&&c^\dag_{j_1}...c^\dag_{j_l}c^\dag_jc^\dag_{j_{l+1}}...c^\dag_{j_n}|0\rangle+...\nonumber\\
&=& \sum_{j<j_1<j_2...<j_n}g_j S(k_1,...,k_n;j_1,...,j_n)c^\dag_jc^\dag_{j_1}...c^\dag_{j_n}|0\rangle+...\nonumber\\
&&+\sum_{j<j_1<j_2...<j_n} g_{j_l} S(k_1,...,k_n;j,j_1,...j_{l-1},j_{l+1},...,j_n)\nonumber\\
&&c^\dag_jc^\dag_{j_1}...c^\dag_{j_n}|0\rangle+...\nonumber\\
&=& \sum_{j<j_1<j_2...<j_n}[\sum^n_{l=1} g_{j_l} S(k_1,...,k_n;j,j_1,...,\underline{j_l},...,j_n)\nonumber\\
&&+ g_j S(k_1,..,k_n;j_1,..,j_n)]\sum_{p_1<...<p_{n+1}}\nonumber\\
&&S^*(p_1,...,p_{n+1};j,j_1,...,j_n)d^\dag_{p_1}...d^\dag_{p_{n+1}}|0\rangle\nonumber\\
&=& \sum_{j_1<j_2<...<j_{n+1}}[\sum^n_{l=1} g_{j_l} S(k_1,...,k_n;j_1,j_2,...,\underline{j_l},...,j_n)]\nonumber\\
&&\sum_{p_1<...<p_{n+1}}S^*(p_1,...,p_{n+1};j_1,j_2,...,j_{n+1})d^\dag_{p_1}...d^\dag_{p_{n+1}}|0\rangle\nonumber\\
&=& \sum_{p_1<...<p_{n+1}}\tilde{f}^*(p_1,...,p_{n+1};k_1,...,k_{n};\{g_j\})d^\dag_{p_1}...d^\dag_{p_{n+1}}|0\rangle\nonumber\\
\end{eqnarray}
Here $(j_1,...,\underline{j_l},...,j_{m+1})$ is the string of length $m$ with the element $j_l$ removed from the sequence $(j_1,...,j_{m+1})$, and the coupling configuration dependent auxiliary $\tilde{f}$-function  is given by Eq. (\ref{ff}) in the main text.
By invoking Eq. (\ref{Sch}), we obtain Eq. (\ref{D1}). The equations of motion for the $B$s (Eq. (\ref{B1})) can be derived similarly.

\section{Appendix C: The ground state structure of the periodic XX chain}
For a periodic XX chain described by $H_B$ with $J=-1$ and $h\ge0$, lowering $h$ from $h_c=1$ to $h=0$ will cause $N/2$ level crossings or parity changing at the following $N/2$ critical fields:
\begin{eqnarray}\label{hcm}
 h_m=-\frac{\cos(m+\frac{1}{2})\frac{\pi}{N}}{\cos\frac{1}{2}\frac{\pi}{N}},~m=\frac{N}{2},\frac{N}{2}+1,...,N-1.
\end{eqnarray}
Note that $h_{N-1}=h_c=1$, so that the region $h\in[0,+\infty)$ is divided into the following intervals:
\begin{eqnarray}\label{hcmh}
&&i): h_m \le h\le h_{m+1},~m=\frac{N}{2},\frac{N}{2}+1,...,N-2,\nonumber\\
&&ii): 1\le h,\nonumber\\
&&iii): 0\le h\le h_{\frac{N}{2}}.
\end{eqnarray}
For fields within interval $i)$ and with $m$ even, the ground state is filled by $m+1$ $d$-fermions
\begin{eqnarray}\label{gd}
|g_m\rangle_e&=&d^\dag_{-m\frac{\pi}{N}}d^\dag_{-(m-2)\frac{\pi}{N}}...d^\dag_0...d^\dag_{(m-2)\frac{\pi}{N}}d^\dag_{m\frac{\pi}{N}}|0\rangle,
\end{eqnarray}
and possesses an energy
\begin{eqnarray}
E^{(e)}_{m}=-(h+1)-2\sum^{\frac{m}{2}}_{l=1}\left(\cos\frac{2\pi l}{N}+h\right).
\end{eqnarray}
Similarly, for odd $m$, the ground state is filled by $m+1$ $c$-fermions
\begin{eqnarray}\label{gc}
|g_m\rangle_o&=&c^\dag_{-m\frac{\pi}{N}}c^\dag_{-(m-2)\frac{\pi}{N}}...c^\dag_{m\frac{\pi}{N}}|0\rangle,
\end{eqnarray}
with energy
\begin{eqnarray}
E^{(o)}_{m}= -2\sum^{\frac{m+1}{2}}_{l=1}\left[\cos\frac{(2l-1)\pi}{N}+h\right].
\end{eqnarray}
For fields within interval $ii)$, the ground state is always the fully polarized state with all spins pointing in the $+\hat{z}$ direction. In the fermionic picture, this state corresponds to the completely occupied state $|g_{N-1}\rangle_o$ which is filled by the $c$-fermions and has an energy $-hN$. Depending on whether $N=4n$ or $N=4n+2$, the ground state for fields within interval $iii)$ will be either $|g_{N/2-1}\rangle_o$ or $|g_{N/2-1}\rangle_e$.
\par For the initial state Eq.(\ref{phi0}) with $|g_{XX}\rangle=|g_m\rangle_e$ (even $m$), $|\phi_I(t)\rangle_{e}$ has a similar form as Eq. (\ref{psiIo}),
\begin{eqnarray}
&&|\phi_I(t)\rangle_{e}=\sum_{k_1<...<k_{m+1}}[a_{\bar{1}}A'(k_1,...,k_{m+1};t)|\bar{1}\rangle\nonumber\\
&&+a_1B'(k_1,...,k_{m+1};t)|1\rangle]\prod^{m+1}_{l=1}d^\dag_{k_l}|0\rangle\nonumber\\
&&+\sum_{k_1<...<k_{m+2}}a_1D'(k_1,...,k_{m+2};t)|\bar{1}\rangle\prod^{m+2}_{l=1}c^\dag_{k_l}|0\rangle\nonumber\\
&&+\sum_{k_1<...<k_{m}}a_{\bar{1}}C'(k_1,...,k_{m};t)|1\rangle\prod^{m}_{l=1}c^\dag_{k_l}|0\rangle.
\end{eqnarray}
The coefficients $A'$, $B'$, $C'$ and $D'$ satisfy a set of equations of motion of the same form as Eq.(\ref{D1}-\ref{B1}) and Eq.(\ref{C}-\ref{A}), except that $m$ is now even. Similar expressions for the $W$-factor also hold, with $A$, $B$, $C$ and $D$ replaced by $A'$, $B'$, $C'$ and $D'$.
\section{Appendix D: Derivation of the short-time behavior of  the decoherence factor $|r(t)|^2$ Eq. (\ref{Gass})}
When the qubit-bath coupling is small compared with the external field $h$ and interaction $J$ between neighboring spins in the bath ($J/g,h/g\gg1$), then standard second order time-dependent perturbation theory can be applied. The time-dependent wavefunction in the interaction picture can be written to second order in the perturbation as
\begin{eqnarray}
|\psi_I(t)\rangle&=&|\phi(0)\rangle+(-i)\int^t_0ds H_I(s)|\phi(0)\rangle+\nonumber\\
&&(-i)^2\int^t_0ds\int^s_0ds' H_I(s)H_I(s')|\phi(0)\rangle
\end{eqnarray}
with $H_I(t)$ and $|\phi(0)\rangle$ given by Eq. (\ref{HI}) and Eq. (\ref{phi0}), respectively. For $m=$even, the initial bath state $|g_{XX}\rangle=\prod^{m+1}_{l=1}c^\dag_{k_l}|0\rangle$, with
\begin{eqnarray}
(k_1,...,k_{m+1})=(-m\frac{\pi}{N},...,m\frac{\pi}{N}).
\end{eqnarray}
Direct calculation gives
\begin{eqnarray}
&&|\phi_I(t)\rangle_o=a_{\bar{1}}\{|\bar{1}\rangle\prod^{m+1}_{l=1}c^\dag_{k_l}|0\rangle- g\sum_{p_1<...<p_m}\nonumber\\
&&  \frac{e^{i(\omega+\sum^m_{l=1}\varepsilon_{p_l}-\sum^{m+1}_{l=1}\varepsilon_{k_l})t}-1}{ \omega+\sum^m_{l=1}\varepsilon_{p_l}-\sum^{m+1}_{l=1}\varepsilon_{k_l} }f(k_1,...,k_{m+1};p_1,...,p_m)\nonumber\\
&&  |1\rangle\prod^m_{l=1}d^\dag_{p_l}|0\rangle+ g^2  \sum_{p_1<...<p_m}\sum_{p'_1<...<p'_{m+1}} \nonumber\\
&&\frac{\frac{  e^{i(\sum^{m+1}_{l=1}(\varepsilon_{p'_l}-\varepsilon_{k_l}))t}-1}{  \sum^{m+1}_{l=1}(\varepsilon_{p'_l}-\varepsilon_{k_l}) } +\frac{e^{-i(\omega+\sum^m_{l=1}\varepsilon_{p_l}-\sum^{m+1}_{l=1}\varepsilon_{p'_l})t}-1}{  \omega+\sum^m_{l=1}\varepsilon_{p_l}-\sum^{m+1}_{l=1}\varepsilon_{p'l} }}{  \omega+\sum^m_{l=1}\varepsilon_{p_l}-\sum^{m+1}_{l=1}\varepsilon_{k_l} }\nonumber\\
&&f(k_1,...,k_{m+1};p_1,...,p_m)  f^*(p'_1,...,p'_{m+1};p_1,...,p_m)\nonumber\\
&& |\bar{1}\rangle\prod^{m+1}_{l=1}c^\dag_{p'_l}|0\rangle\}\nonumber\\
&&+a_1\{|1\rangle\prod^{m+1}_{l=1}c^\dag_{k_l}|0\rangle + g \sum_{p_1<...<p_{m+2}}\nonumber\\
&&  \frac{e^{i (\omega-\sum^{m+2}_{l=1}\varepsilon_{p_l}+\sum^{m+1}_{l=1}\varepsilon_{k_l}) t} -1 }{ \omega-\sum^{m+2}_{l=1}\varepsilon_{p_l}+\sum^{m+1}_{l=1}\varepsilon_{k_l} } f^*(p_1,...,p_{m+2};k_1,...,k_{m+1}) \nonumber\\
&&|\bar{1}\rangle\prod^{m+2}_{l=1}d^\dag_{p_l}|0\rangle - g^2  \sum_{p_1<...<p_{m+2}}\sum_{p'_1<...<p'_{m+1}}\nonumber\\
 &&\frac{   \frac{e^{i\sum^{m+1}_{l=1}(\varepsilon_{p'_l}-\varepsilon_{k_l})t}-1}{ \sum^{m+1}_{l=1}(\varepsilon_{p'_l}-\varepsilon_{k_l})} -\frac{e^{i(\omega-\sum^{m+2}_{l=1}\varepsilon_{p_l}+\sum^{m+1}_{l=1}\varepsilon_{p'_l})t}-1}{ \omega-\sum^{m+2}_{l=1}\varepsilon_{p_l}+\sum^{m+1}_{l=1}\varepsilon_{p'_l} }  }{  \omega-\sum^{m+2}_{l=1}\varepsilon_{p_l}+\sum^{m+1}_{l=1}\varepsilon_{k_l} } \nonumber\\
&&f^*(p_1,...,p_{m+2};k_1,...,k_{m+1})  f(p_1,...,p_{m+2};p'_1,...,p'_{m+1})  \nonumber\\
&&|1\rangle\prod^{m+1}_{l=1}c^\dag_{p'_l}|0\rangle\}.
\end{eqnarray}
Comparing with Eq. (\ref{psiIo}) in the main text, we have
\begin{eqnarray}
&&A(k_1,...,k_{m+1};t)=1+g^2\sum_{p_1<...<p_m}\nonumber\\
&&\frac{ e^{-i(\omega+\sum^m_{l=1}\varepsilon_{p_l}-\sum^{m+1}_{l=1}\varepsilon_{k_l})t}-1}{ (\omega+\sum^m_{l=1}\varepsilon_{p_l}-\sum^{m+1}_{l=1}\varepsilon_{k_l})^2}|f(k_1,...,k_{m+1};p_1,...,p_m)|^2,\nonumber\\
&&B(k_1,...,k_{m+1};t)=1+g^2 \sum_{p_1<...<p_{m+2}}\nonumber\\
&& \frac{e^{i(\omega+\sum^{m+1}_{l=1}\varepsilon_{k_l}-\sum^{m+2}_{l=1}\varepsilon_{p_l})t}-1}{ (\omega+\sum^{m+1}_{l=1}\varepsilon_{k_l}-\sum^{m+2}_{l=1}\varepsilon_{p_l})^2}
|f(p_1,...,p_{m+2};k_1,...,k_{m+1})|^2,\nonumber\\
\end{eqnarray}
and
\begin{eqnarray}
&& A(p'_1,...,p'_{m+1};t)=O(g^2),\nonumber\\
&&  B(p'_1,...,p'_{m+1};t)=O(g^2),
\end{eqnarray}
 for $(p'_1,...,p'_{m+1})\neq(k_1,...,k_{m+1})$. From Eq. (\ref{W}), we finally obtain
 \begin{eqnarray}
|r(t)|^2 &=& 1+2g^2 \sum_{p_1<...<p_m} |f(k_1,...,k_{m+1};p_1,...,p_m)|^2\nonumber\\
&&\frac{ \cos (\omega-\sum^{m+1}_{l=1}\varepsilon_{k_l}+\sum^{m }_{l=1}\varepsilon_{p_l})t-1}{ (\omega+\sum^m_{l=1}\varepsilon_{p_l}-\sum^{m+1}_{l=1}\varepsilon_{k_l})^2}\nonumber\\
 &&+2g^2  \sum_{p_1<...<p_{m+2}} |f(p_1,...,p_{m+2};k_1,...,k_{m+1})|^2\nonumber\\
 &&\frac{\cos(\omega+\sum^{m+1}_{l=1}\varepsilon_{k_l}-\sum^{m+2}_{l=1}\varepsilon_{p_l})t-1}{ (\omega+\sum^{m+1}_{l=1}\varepsilon_{k_l}-\sum^{m+2}_{l=1}\varepsilon_{p_l})^2}.
\end{eqnarray}
For short times $ (\omega-\sum^{m+1}_{l=1}\varepsilon_{k_l}+\sum^{m }_{l=1}\varepsilon_{p_l})t,(\omega+\sum^{m+1}_{l=1}\varepsilon_{k_l}-\sum^{m+2}_{l=1}\varepsilon_{p_l})t\ll1$, we have
 \begin{eqnarray}
|r(t)|^2 &\approx&1-\alpha(gt)^2\approx e^{-\alpha(gt)^2},
\end{eqnarray}
 with $\alpha$ given by Eq. (\ref{alpha}).

\end{document}